\documentclass[useAMS,usenatbib]{mnras}

\usepackage{url,times,hyperref,graphicx,amsmath,amsfonts,amssymb,aas_macros,color,epsfig,epstopdf}

\newcommand{\mhalo}{M$_{\rm halo}$}
\newcommand{\mstar}{M$^{\star}$}

\newcommand{\msun}{M$_\odot$}

\usepackage{float}

\newcommand{\hMpc}{{\ifmmode{h^{-1}{\rm Mpc}}\else{$h^{-1}$Mpc}\fi}}
\newcommand{\hkpc}{{\ifmmode{h^{-1}{\rm kpc}}\else{$h^{-1}$kpc}\fi}}
\newcommand{\hMsun}{{\ifmmode{h^{-1}{\rm {M_{\odot}}}}\else{$h^{-1}{\rm{M_{\odot}}}$}\fi}}
\newcommand{\ltsima}{$\; \buildrel < \over \sim \;$}
\newcommand{\gtsima}{$\; \buildrel > \over \sim \;$}
\newcommand{\lsim}{\lower.5ex\hbox{\ltsima}}
\newcommand{\gsim}{\lower.5ex\hbox{\gtsima}}

\def\lesssim{\mathrel{\hbox{\rlap{\hbox{\lower4pt\hbox{$\sim$}}}\hbox{$<$}}}}
\def\gtrsim{\mathrel{\hbox{\rlap{\hbox{\lower4pt\hbox{$\sim$}}}\hbox{$>$}}}}

\newcommand{\Sec}[1]{Section~\ref{#1}}
\newcommand{\Eq}[1]{Eq.~(\ref{#1})}
\newcommand{\Fig}[1]{Fig.~\ref{#1}}
\newcommand{\beq}{\begin{equation}}
\newcommand{\eeq}{\end{equation}}
\def\beqa{\begin{eqnarray}}
\def\eeqa{\end{eqnarray}}
\def\hMpc{$h^{-1}\,{\rm Mpc}$}
\def\hkpc{$h^{-1}\,{\rm kpc}$}

\def\head{
 \vbox to 0pt{\vss
                   \hbox to 0pt{\hskip 440pt\rm LA-UR-10-07069\hss}
                  \vskip 25pt}}

\title[Black holes  in  SIDM galaxies]
{A rumble in the dark: signatures of self-interacting dark matter in Super-Massive Black Hole dynamics and galaxy density profiles}
\author[Di Cintio]
       {Arianna Di Cintio$^{1,2}$
\thanks{E-mail: arianna.dicintio@dark-cosmology.dk, adicintio@aip.de} \thanks{DARK-Carlsberg and Karl Schwarzschild fellow}, Michael Tremmel$^{3}$, Fabio Governato$^{3}$, Andrew Pontzen$^4$,\newauthor  Jes\'{u}s Zavala$^{5}$, Alexander Bastidas Fry$^3$, Alyson Brooks$^{6}$ \& Mark Vogelsberger$^7$ \thanks{Alfred P. Sloan Fellow}\\
$^{1}$Dark Cosmology Centre, Niels Bohr Institute, University of Copenhagen, Juliane Maries Vej 30, DK-2100 Copenhagen, Denmark\\
$^2$ Leibniz Institute for Astrophysics Potsdam (AIP), An der Sternwarte 16, D-14482 Potsdam, Germany\\
$^{3}$Astronomy Department, University of Washington, PO Box 351580, Seattle, WA, 98195-1580, USA\\
$^4$Department of Physics and Astronomy, University College London, London WC1E 6BT, UK\\
$^5$ Center for Astrophysics and Cosmology, Science Institute, University of Iceland, Dunhagi 5, 107 Reykjavik, Iceland\\
$^6$ Department of Physics \& Astronomy, Rutgers, The State University of New Jersey, 136 Frelinghuysen Rd, Piscataway, NJ 08854, USA\\
$^7$ Department of Physics, Kavli Institute for Astrophysics and Space Research, MIT, Cambridge, MA 02139, USA}
\begin{document}

\date{12 January 2017}

\pagerange{\pageref{firstpage}--\pageref{lastpage}} \pubyear{2010}

\maketitle

\label{firstpage}

\begin{abstract}
We explore for the first time  the effect of self-interacting dark matter (SIDM) on the dark matter (DM) and baryonic distribution in massive  galaxies formed in  hydrodynamical cosmological simulations, including  explicit baryonic physics treatment.
 A novel implementation of Super-Massive Black Hole (SMBH)  formation and evolution is used, as in \citet{tremmel15,tremmel16}, allowing  to explicitly follow SMBH dynamics at the center of galaxies.
A high SIDM constant cross-section is chosen, $\sigma$=10 $\rm cm^2/gr$, to amplify differences from  CDM models.
Milky Way-like galaxies form a shallower DM density profile in SIDM than they do in CDM, with differences already at 20 kpc scales. This demonstrates that even for the most massive spirals the effect of SIDM dominates over the adiabatic contraction due to baryons.  
\noindent Strikingly, the dynamics of SMBHs differs in the SIDM and reference CDM case. 
SMBHs in massive spirals have sunk to the centre of their host galaxy in both the SIDM and CDM run, while in less massive galaxies about 80$\%$ of the SMBH population is off-centered  in the SIDM case, as opposed to the CDM case in which $\sim$90$\%$ of SMBHs have reached their host's centre. SMBHs are found as far as $\sim$9 kpc away from the centre of their host SIDM galaxy.
This difference is due to the increased dynamical friction timescale caused by the lower DM density in  SIDM galaxies compared to CDM, resulting in   \textit{core stalling}.

\noindent This pilot work highlights the importance of  simulating in a full hydrodynamical context different DM models combined to  SMBH physics to study their influence on galaxy formation.
\end{abstract}

\noindent
\begin{keywords}
 galaxies: evolution - formation - haloes cosmology: theory - dark matter
 \end{keywords}

\vspace{-1cm}\section{Introduction} \label{sec:introduction}
Self-interacting  dark  matter (SIDM), originally introduced over a decade ago by \citet{spergel00} as a heuristic model to solve the problem of  observed shallow dark matter (DM) profiles in galaxies,  is also the simplest case of non-standard DM structure formation  models with `dark sector' interactions. SIDM has recently captured an increasing interest within the community.
The collisional, self-scattering particles can create cores of dark matter within galaxies by  transferring mass  from the dense central regions of DM haloes, where  the probability of collisions is higher, toward the halo outskirts \citep{balberg02,colin02,koda11}.
This process represents a viable solution to the so-called \textit{core-cusp}  problem \citep{moore94,oh08,walker11,adams14}.

The rate of collisions, determined by the cross-section per unit mass $\sigma/m$ (from now on simply $\sigma$) is constrained from several astrophysical observations, such as the necessity of forming cores in very faint galaxies without evaporating the satellites of Milky Way-sized haloes or the galaxies in clusters, maintaining the ellipsoidal shape of haloes and clusters and avoiding the gravothermal catastrophe \citep{Gnedin01,firmani01,peter13,robertson16}.
Several authors have successfully run SIDM simulations placing  further constraints on cross-sections that are constant across all interaction velocities, and found  that the relevant range to impact galaxy evolution and avoid upper limits lies between  0.1$<$$\sigma/\rm (cm^2/gr)$$<$1 \citep{peter13,vogelsberger12,Vogel13,rocha13,zavala13,ethosI,ethosII}.
A velocity-dependent cross section could, however, ease the constraints on $\sigma$ by allowing the dark matter to behave as a collisional fluid in dwarfs, and as a collisionless one at clusters scales \citep{yoshida00,colin02,elbert16}.
These simple predictions need, however, to be evaluated in the presence of baryonic processes such as supernovae (SN) driven outflows.  SN-driven winds  remove  the excess  of low angular momentum gas  and explain the formation of bulgeless galaxies \citep{governato10}, which no alternative DM model solves. Outflows also predict, in agreement with  observations, the  formation of shallow DM profiles at the centre of galaxies \citep[e.g.][]{governato12,Pontzen12,DiCintio2014a,DiCintio2014b,brooks14,Pontzen14,onorbe15,tollet16}.

Comparing the predictions of  SIDM and cold dark matter (CDM)  models once coupled to baryon physics has been recently explored in low mass galaxies \citep{vogelsberger14,bastidas15}. \citet{vogelsberger14} show that the stellar core in simulated SIDM dwarfs is closely related to the dark matter core radius generated by self-interactions. In \citet{bastidas15} by choosing a  relatively  large  cross-section of 2 cm$^2$/gr and by including the mechanisms able to create a core through supernovae feedback and bursty star formation \citep{Pontzen14}, the authors showed that the DM profiles and star formation (SF) histories of  dwarf galaxies in SIDM simulations do not essentially differ from CDM ones, both being in agreement with observational results.

Attempts to calculate analytically the response of SIDM particles in the presence of baryons have been made by \citet{kaplinghat14}: following the scaling relation for SIDM presented in \citet{rocha13}, they initially showed that by using a cross-section of $\sigma$=0.56 $\rm cm^2/gr$ 
the deviations in density profile of a Milky-Way halo,  due to self-interactions, are expected at radii $\lesssim$10 kpc. 
\citet{kaplinghat14}  claimed, however, that such analytic prediction holds for SIDM-only simulations, since the  presence  of  baryons  changes  the  SIDM density profile by decreasing the core radius and increasing the core density: the expected core size in a  Milky Way galaxy, for $\sigma$=0.56 $\rm cm^2/gr$ and accounting for  the gravitational potential of both baryons and dark matter, would be around  0.3  kpc, more than an order of magnitude smaller than the core size from SIDM-only simulations. Such analytic treatment has been  shown to be in agreement with results from idealised simulations of SIDM with baryons, by  \citet{elbert16}. If proven correct, this would  imply that the adiabatic contraction effect due to  baryons may loosen the constraints on SIDM cross sections.

This  prediction deserves however further investigation, since it heavily relies on analytic models and  idealised simulations,  ignoring the complexity  of galaxy formation processes such as violent relaxation caused by mergers and outflows created by both SN and Super-Massive black holes (SMBHs) that regulate  SF efficiency in galaxies.

In this paper we explore for the first time the effect, on galaxies of different masses, of a large constant cross section in SIDM simulations, including baryonic physics and SMBHs, and we compare the results against a standard CDM+baryons run. We further study and highlight the differences in SMBH dynamics in the two cosmologies, by explicitly following SMBH orbital decay:
this is a unique capability of our runs and represents a step forward compared to previous work in the field.
We run a box of 8 Mpc in side up to z=0.5, with the most massive galaxy  being a Milky Way analogue. Both runs include a novel parameterisation of SMBH physics following \citet{tremmel15} and \citet{tremmel16}, in which SMBHs are allowed to form in dense pristine gas regions and their orbits can be followed as they sink toward the galaxy center due to  dynamical friction forces \citep{chandrasekhar43,BT08}. This approach is a significant improvement over previous `advection' schemes that  force SMBHs at the  galaxy center during merger events or satellite accretion \citep{dimatteo05,sijacki07}, resulting in unrealistic timescales for SMBH orbital decays.
In order to highlight differences between CDM and SIDM, we used a cross section of $\sigma$=10 $\rm cm^2/gr$, which is allowed at the scale of the Milky-Way, and it is in agreement with the upper limits derived by \citet{kaplinghat15} using rotation curves of low surface brightness galaxies.

This manuscript is organised as follows: in \Sec{sec:model} we show the characteristics of the simulated galaxies, including a full description of self-interactions, SMBH and stellar physics implementations; in \Sec{sec:results} we discuss the main results, focusing on DM density profiles, star formation histories and SMBH properties in massive (\Sec{sec:massive}) and intermediate mass galaxies (\Sec{sec:dwarfs}) and on the global properties of the SMBH population in SIDM and CDM cosmologies (\Sec{sec:SMBHs} and \Sec{sec:SMBHs_dyn}); we conclude in \Sec{sec:concl}.

\vspace{-.6cm}\section{Simulations} \label{sec:model}
We run  hydrodynamical simulations of  the formation of galaxies in a full cosmological context, within a box of 8 Mpc in side, employing  cosmological parameters  from the latest Planck results in a $\Lambda$ dominated  universe ($\Omega_0$=0.3, $\Lambda$=0.7, h=0.67, $\sigma_8$=0.83, \citealt{planck13}) and  following the evolution of structure formation until z=0.5. 
We used two underlying models for dark matter, a Cold Dark Matter (CDM) and  a Self-Interacting (SIDM) one, with the same set of initial conditions. We employ a constant cross-section of $\sigma$=10 $\rm cm^2/gr$ for the SIDM model. The simulations are run using the new N-body + SPH code ChaNGa\footnote{www-hpcc.astro.washington.edu/tools/changa.html} \citep{menon15}, which is an improved version of the code Gasoline and  includes several standard modules such as a cosmic UV background, star formation, and blastwave feedback from supernovae \citep{wadsley04,wadsley08,stinson06}. 
The SPH implementation also includes thermal diffusion
\citep{shen10} and eliminates artificial gas surface tension
through the use of a geometric mean density in the SPH
force expression \citep{menon15,governato15}. This update better simulates shearing
flows with Kelvin-Helmholtz instabilities.

The simulations are run with a  gravitational force spline softening length of 350 pc.
The dark matter is oversampled such that we simulate $\sim$3 times more dark matter particles than gas particles, resulting in a dark matter particle mass of $3.4\times10^5$\msun\  and gas
particle mass of $2.1\times10^5$\msun. This methodology  decreases
numerical noise and improves SMBH dynamics \citep{tremmel15}. The  halo masses are defined as the mass of a sphere containing $\rm \Delta_{vir}$ times the critical matter density of the Universe, such that \mhalo=$\rm 4\pi R_{vir}^3 \Delta_{vir}\rho_{crit}/3$ and $\rm \Delta_{vir}$ depends on the chosen cosmology \citep{Bryan98}. For a Planck cosmology, $\rm \Delta_{vir}$$\sim$ 100.  The  haloes are identified using the AHF\footnote{http://popia.ft.uam.es/AHF/Download.html} halo finder \citep{Knollmann09} and analysed with the \textit{pynbody}\footnote{https://pynbody.github.io/pynbody/installation.html} package \citep{Pontzen13} and the TANGOS database (Pontzen  et  al.  in  prep). 

Overall, the feedback mechanisms implemented in the simulations are able to create  galaxies with the expected stellar mass per each halo mass, despite of the underlying dark matter model, SIDM or CDM, as  shown in \citet{tremmel16} and \citet{bastidas15}.

\subsection{Black Hole dynamics and accretion: a new model}

SMBH physics has been implemented following the novel approach of \citet{tremmel15,tremmel16}.

The SMBH seed formation is connected to the physical state of the gas in the simulation at high redshift, without any a priori assumptions regarding halo occupation fraction: this allows to naturally populate galaxies of different masses with SMBHs. SMBHs form in the early Universe from dense, pristine low metallicity (Z$<3\times10^{-4}$) gas with densities 15 times higher than the star formation threshold ($3m_p/cm^3$)  with temperatures between 9500 K and 10000 K.
Seed SMBH formation is then limited to the highest density peaks in the early Universe with high Jeans masses. 
This technique forms SMBH seeds within the first billion years of the simulation, allowing us to follow their dynamics throughout the assembly of the host halo, even for small haloes.

An important improvement, which made this study possible for the first time, regards the treatment of dynamical friction. The gravitational wake of a massive body moving in the extended potential of a medium will causes the orbit of  SMBHs to decay towards the center of massive galaxies \citep{chandrasekhar43,BT08}. Previous authors have provided  analytic expression to compute the dynamical friction timescales  $\rm t_{df}$ of rigid bodies merging at the centre of galaxies \citep{taffoni03,boylan07}, showing that it can easily exceed several Gyrs, making previous `advection' techniques inappropriate to realistically model such a significant timescale for sinking SMBHs \citep{dimatteo05,sijacki07}. Indeed, in `advection' models the SMBHs are  re-positioned and forced at the  galaxy center during merger events or satellite accretion, resulting in unrealistic timescales for SMBH orbital decays.

In this work we instead use a novel prescription firstly introduced in  \citet{tremmel15,tremmel16}, which includes a sub-grid approach for modelling unresolved dynamical friction on scales smaller than the gravitational softening length, adding  a force correction to the SMBH acceleration.
The dynamical friction force is:
\begin{equation}
 \vec{\rm F}_{\rm df}=\rm -4\pi G^2ln(\Lambda)M\rho_{host}(<v_{orb})/v_{orb}^2
\end{equation}
where M is the mass of the object with orbital velocity $\rm v_{orb}$, $\rm \rho_{host}$ is the density of host background particles with velocities less than the orbital velocity and $\rm \Lambda$ is the usual Coulomb logarithm.
Such  acceleration is added to the SMBHs current acceleration, and integrated in the following time step. 
The resulting sinking timescale $\rm t_{df}$ will be therefore dependent on the density of the surrounding galaxy, and on the mass and the velocity of the SMBH itself. This technique has been shown to produce realistically sinking SMBHs \citep{tremmel15}.
We are consequently able to resolve the dynamics of SMBHs during and after galaxy mergers down to sub-kpc scales, with important implications for  the dynamics of SMBHs in galaxies with different underlying densities.

Once formed, the seed mass is set to $10^6$\msun. 
SMBHs will then accrete gas according to a modified Bondi-Hoyle formula that accounts for the rotational support of gas \citep{tremmel15,tremmel16}; seeds that exist in unfavorable environments, such as dwarf galaxies, will naturally have limited growth over a Hubble time.
The energy from accretion is then isotropically transferred to nearby gas particles with a technique similar
to the blastwave supernovae feedback \citep{stinson06}: cooling is turned off for the gas particles immediately surrounding the SMBH,  resembling the continuous transfer of energy during each SMBH timestep.
The amount of energy coupled
to surrounding gas particles is given by $\rm E=\epsilon_r\epsilon_f\dot{M}c^2dt$, where $\dot{M}$ is the accretion rate, $\epsilon_r$ the radiative efficiency and $\epsilon_f$ the efficiency of energy that couples to gas (see \citealt{tremmel16} for details about the calibration of these parameters).
Because SMBH growth depends on the host galaxy mass, SMBH feedback is able to preferentially limit
the growth of  massive galaxies, while not quenching the
star formation in low mass haloes.

\subsection{Star Formation Recipes}
The parameters associated with stellar physics have been tuned to result in the
most realistic galaxies possible at z=0, within the implementation of our subgrid model \citep{tremmel16}. Star formation in ChaNGa is regulated through a series of sub-grid prescriptions that parametrize unresolved physics into several free parameters.
Stars are formed stochastically in cold (T$<$$10^4$K) gas
that exceeds a  density of $n_{th}$=0.2$m_p/cm^3$.  Supernova feedback is implemented using the \citet{stinson06} blastwave formalism, depositing E$_{SN}$=10$^{51}$ erg into the surrounding interstellar medium at the end of the lifetime of stars more massive than 8\msun. A Kroupa IMF \citep{kroupa02} is employed to compute the number of stars that will end as SN.
Due to the relatively low resolution of the simulations, we do not expect significant core formation from  baryonic feedback, which needs a $n_{th}$ of at least  $10m_p/cm^3$ to effectively transfer energy to the DM (see \citealt{governato10,Pontzen12} for details). In this work we are focusing on a setup that does not foreseen SN-driven DM core formation to avoid, for now, a more complicated scenario.

\subsection{Self-interactions implementation}
The SIDM model is implemented in a  similar way as in \citet{bastidas15}, and we refer to their work for a comprehensive explanation of the methodology and further details. In this work, we employ  a high constant cross-section of $\sigma$=10 $\rm cm^2/gr$, which is allowed at scales of the Milky Way \citep{kaplinghat15}, in order to maximise the effects of self-interactions on  DM haloes at every mass. 
SIDM interactions are modelled  assuming that each simulated DM particle
represents a  phase-space density patch and the probability of collisions is derived from the collision term in the Boltzmann equation. When a particle collision is detected, the particles are isotropically and elastically scattered, explicitly conserving energy. These interactions are more common in the inner region of the  halo where the density is higher.
The collisions between dark matter particles will then result in a net transfer of mass outwards from the dense central regions of DM haloes, over cosmic time scales, in a process that creates large cores and more spherical haloes with respect to the CDM case \citep{burkert00,spergel00}.

Following the analytic model of  \citet{rocha13} and \citet{bastidas15} (see also \citealt{dooley16} for an updated model) we can separate the dark matter halo into two regions, delimited by a characteristic radius  r, imposing that at such radius the average number of scattering per particles, $\Gamma$,  for the entire life of a galaxy, t, is  unity:
\begin{equation}
\rm \Gamma\approx1\approx t\times \rho(r) \times v(r) \times \sigma
\label{rateSI}
\end{equation}
where v(r) is the velocity dispersion of dark matter at radius r and $\rho(r)$ is the dark matter density at the same radius and t$\sim$10 Gyrs for a Milky Way galaxy.  

\begin{figure*}
\hspace{-.3cm}\includegraphics[width=3.6in,height=2.5in]{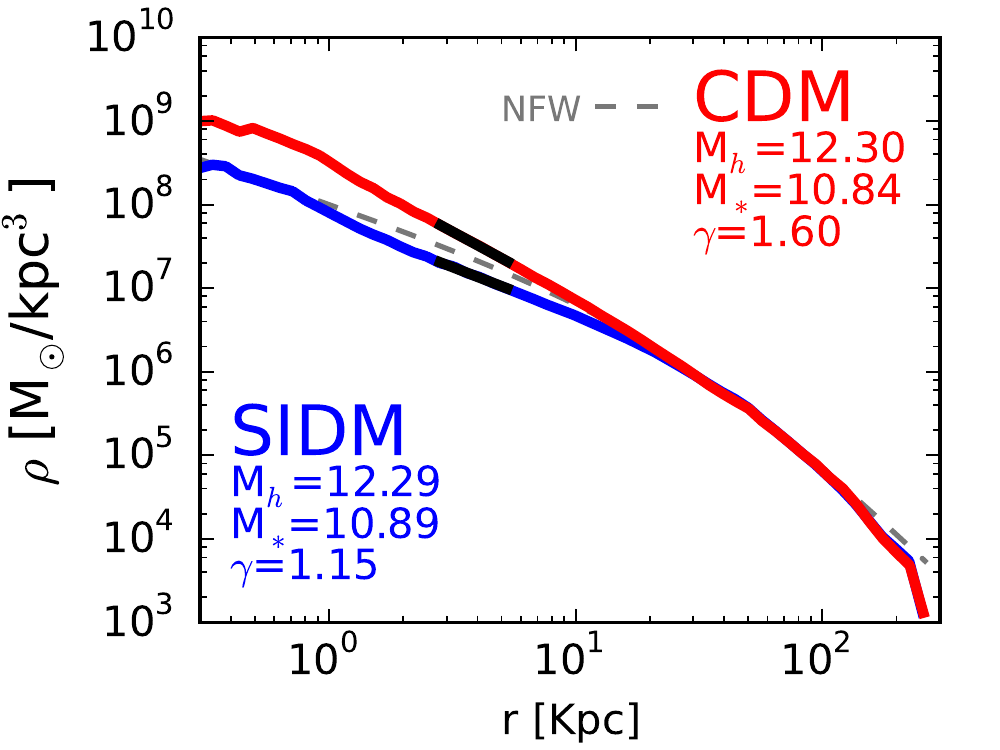}\includegraphics[width=3.6in,height=2.5in]{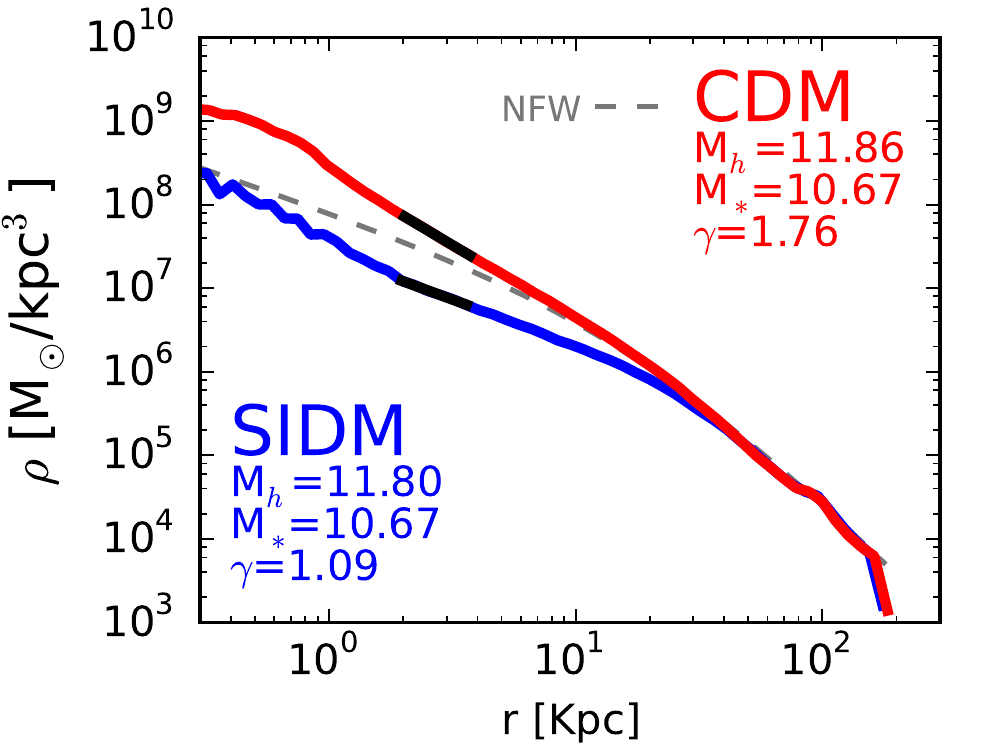}
\caption{Dark matter haloes density profiles for the two most massive, Milky Way-sized galaxies in our simulations. SIDM results are plotted in blue and CDM ones in red. The halo and stellar mass in the SIDM and CDM run are indicated, together with the DM inner slope $\gamma$ computed between 1-2$\%$R$_{\rm vir}$. An unmodified NFW profile of similar halo mass is shown as dashed grey line in each panel. The effect of self-interactions is dominant over the one of adiabatic contraction even in such large, baryon dominated galaxies, as it can be appreciated in  the lower density of  SIDM haloes compared to  CDM ones.}
\label{fig:DM-density} 
  \end{figure*}

At radii larger than the characteristic radius, the  scattering   occurs less  than  once  per  particle,  on  average, and we  expect  the  DM density profile to be unaffected by the collisions, maintaining the prediction  typical of collisionless CDM haloes, i.e. the \citet{navarro96} (NFW) profile.  Within the characteristic radius, instead, scattering from self-interactions happens more than once per particle during a time interval of t: here is where  we expect DM particles to modify the inner DM profile.

We can use \Eq{rateSI} and the scaling of $\rm \rho v\approx (t\sigma)^{-1}$ as a function of $\rm r/r_s$, as in Fig. 7 of \citet{rocha13}, where $\rm r_s$ denotes the scale radius of an initial NFW halo, to estimate the radius within which collisions are important, as a function of  cross-section.
Note that   the core radius due to self-interactions will then be  a fraction of this special radius. 
With our value of $\sigma$=10 $\rm cm^2/gr$, and over a timescale of 10 Gyrs, we expect  modifications in the density profile of a  Milky Way like halo   to happen already at a radius $\rm r/r_s\sim2$ which, for a typical $\rm r_s$=20kpc, leads to physical radii of 30-40 kpc. 

As pointed out in \citet{bastidas15} at the lowest, dwarf scales, not every SIDM model will necessarily form significant DM cores: this is because, according to \Eq{rateSI}, low halo velocities and low central densities may result in a timescale for collisions comparable to or longer than the lifetime of the halo itself, which will thus maintain the DM cusp. In \citet{zavala13}, for example, it has been shown that a cross-section of $\sigma$=0.1 $\rm cm^2/gr$ will still produce a populations of satellites with higher-than-observed densities compared to the MW dwarf spheroidals, and that a $\sigma$$\geqslant$1 $\rm cm^2/gr$ is needed to solve the discrepancy.
With our choice of $\sigma$=10 $\rm cm^2/gr$ and for a timescale of 10 Gyrs, we should expect cores of the order of the scale radius even for  galaxies with \mhalo$\sim10^{10.0-10.5}$\msun.

\section{Results} \label{sec:results}

We study the  dark matter density profiles, star formation histories and SMBH dynamics in galaxies of different masses within the  SIDM and CDM cosmology. We focus on haloes more massive than \mhalo=$10^{10}$\msun\ since galaxies of lower halo masses in both SIDM and CDM have extensively been studied elsewhere \citep{vogelsberger12,rocha13,zavala13,vogelsberger14,bastidas15}.

SMBHs release energy in the surrounding medium, having an effect on both star formation history \citep{tremmel16} and possibly on the dark matter distribution within galaxies. Self-interactions will cause the inner haloes to have a hot core, indicative of heat transport from the outskirts inwards \citep{balberg02,colin02,koda11}: with a cross-section of $\sigma$=10 $\rm cm^2/gr$  we expect to maximise the effect of core formation even in small, \mhalo$\sim$$10^{10}$\msun\ galaxies. 
 Moreover,  the effect of gas inflows during the process of galaxy formation is to slowly drag dark matter toward the galaxy centre,  in a process known as adiabatic contraction \citep{Blumenthal86,Gnedin11}, which  renders the centre of dark matter haloes  more concentrated than what is found in N-body only simulations. The adiabatic contraction process is particularly important in massive galaxies, where the efficiency of star formation is the highest.
Finally, note that due to limited resolution core formation from baryons \citep{Pontzen12} will be a lower limit, i.e. larger cores are expected in higher resolution runs, which we plan to address in future work.

The goal of this section is to probe the respective contribution of the several above mentioned competing effects, which may modify  the dark matter distribution within galaxies as well as affect their star formation histories (SFHs) and their SMBH dynamics.

\subsection{Massive, L$_{\star}$ galaxies} \label{sec:massive}
In \Fig{fig:DM-density} we show the DM density profiles of the two most massive, Milky Way-sized haloes within the simulated volume. SIDM results are shown in blue and CDM ones in red. In both cosmologies, the most massive galaxy (left panel) has a mass of  \mhalo$\sim10^{12.3}$\msun\  while the second most massive one (right panel) has \mhalo$\sim10^{11.8}$\msun, both values being within current constraints for the MW mass \citep[e.g.][]{Xue08,Cautun14}.
The halo  and stellar mass are indicated for each galaxy, together with the DM logarithmic density slope $\gamma$ measured at 1-2$\%\rm R_{vir}$. A reference NFW halo of same  mass is shown as  dashed grey line.
\begin{figure}
\hspace{-.3cm}\includegraphics[width=3.6in,height=4.5in]{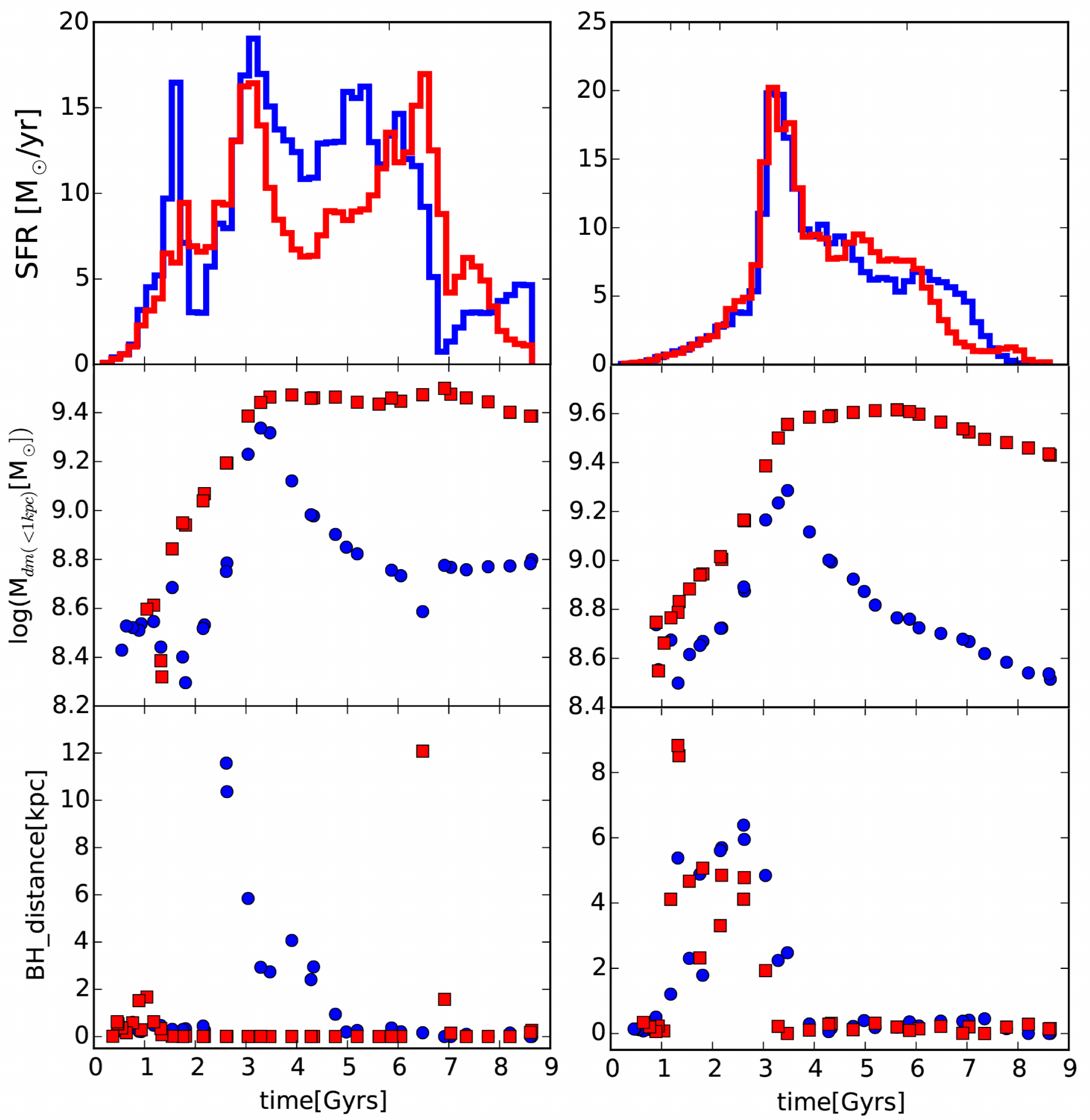}
\caption{From top to bottom: SFHs, evolution of central DM mass and evolution of SMBH distance from the galaxy centre, for the same galaxies 
shown in \Fig{fig:DM-density}. SIDM results are indicated as blue lines and circles, while the CDM ones as red lines and squares. Only the most massive SMBH within each galaxy is followed back in time. See text for  details.}
\label{fig:DM-SFH} 
 \end{figure}

The highest masses, Milky Way-like galaxies, form a shallower DM density profile in SIDM than they do in CDM.
The shallower density profile of dark matter in the SIDM galaxies can be entirely attributed to the effect of self-interactions which, by transferring heat from the outskirts toward inside, are able to generate a lower density already at scales of 20-30 kpc. We stress here that we do not find a dark matter core with zero slope out to 20-30 kpc, but rather observe a decreasing of the DM density out to this radius.
The density profile of the two most massive CDM galaxies, instead, is contracted with respect to expectations from a  CDM-only NFW halo, with a logarithmic inner slope measured at 1-2$\%$$\rm R_{vir}$ of $\gamma$=1.6-1.8: in absence of self-interactions, the contribution of adiabatic contraction is thus dominant  in the CDM run.

Our study  shows that self-interactions dominate over adiabatic contraction in massive galaxies. Using \Eq{rateSI} and for a cross-section of $\sigma$=10 $\rm cm^2/gr$, the self-interactions are expected to modify and lower the DM density profile of MW objects already at radii $\sim$30 kpc. In the  hydrodynamical SIDM run we observe indeed a lowering of the SIDM profile at a similar radius as we would expect in the SIDM-only case, rather than at a much smaller radii as suggested by \citet{kaplinghat14} and \citet{elbert16}. 
\citet{kaplinghat14} predicted that the dark matter core sizes  in baryon dominated galaxies will be negligible, with the core size of MW galaxies expected to be more than an order of magnitude smaller than the core size from SIDM-only simulations, for a cross section of $\sigma$=0.56 $\rm cm^2/gr$.  Similarly, a recent study by \citet{elbert16}  indicated that  when the stellar gravitational potential  dominates  the centre of galaxies,  SIDM haloes can be as dense as  CDM ones. 
Our results differ from  the results of \citet{kaplinghat14}  and \citet{elbert16} for two reasons: on one side, they used a much smaller cross section than we do ($\sigma$=0.56 $\rm cm^2/gr$), and on another side, their work relied  on analytic equilibrium models and  idealised simulations, respectively, rather than on detailed hydrodynamical cosmological simulations able to capture the complexity of galaxy formation, as in our case. 

In \citet{elbert16}, the authors used N-body simulations and analytic galaxy potentials designed to grow linearly in time, with different scale lengths, in order to explore the halo back-reaction to the growth of DM and disk component. While the \citet{kaplinghat14} model is devoted to treat contraction in SIDM halos due to baryonic potential in an analytic manner. The main problem with these approaches is that they both ignore the effect of feedback on DM haloes, which is very relevant especially when combined with the one of self-interactions: while it is true that stellar feedback alone has not been found to affect the final DM distribution in massive, MW like galaxies  (see  \citealt{DiCintio2014a} and references therein), no simulations has explored so far the effect of stellar and BH feedback acting on initially different DM profiles, such as the SIDM one studied in this work.
Feedback can affect the central distribution of galaxies by powering massive outflows of gas in the interstellar medium on  timescale  comparable to the crossing time of  DM particles, which creates a way for transferring energy from gas to DM. In particular, the effects of BH and stellar feedback may be different when the underlying density and gravitational potential are lower than the reference CDM scenario, as in the case of self-interacting galaxies. There are indeed evidences that, in the presence of self-interactions, outflows have the ability to extend further out in the galaxy, and therefore to have a stronger impact on the DM distribution \citep{vogelsberger14}. Given the non-adiabatic nature of such processes, it is important to study them with hydrodynamical simulations.

Further, we verified in our runs that the baryonic  potential  dominates  the centre of the  massive galaxies shown in \Fig{fig:DM-density} and yet the SIDM halo shows a lower DM density than the CDM case. Specifically, we found the same amount of baryonic mass ($\sim$2.2-2.5$\times10^{10}$\msun) within the inner 1 kpc of the SIDM and CDM galaxies, but a much lower content in DM in the SIDM case. The baryonic disk dominates the matter content out to radii as large as 8-9 kpc for the CDM case, and even larger for the SIDM galaxies, given the reduced contribution from the DM component.
This means that even when the central potential of massive galaxies is dominated by baryons, in a similar way in CDM and SIDM galaxies, their final DM density profile does not necessarily follow such potential and it can  be strongly modified by the  combined effect of self-interactions, stellar and BH feedback, highlighting the importance of simulating  galaxies with full hydrodynamics.

In \Fig{fig:DM-SFH} we show, from top to bottom, the star formation histories, temporal evolution of DM mass within 1 kpc and time evolution of the distance of the most massive SMBH from the galactic centre, for the two Milky Way-sized haloes of \Fig{fig:DM-density}.
SIDM results are shown in blue and circled points while CDM ones are shown in red and squares symbols.
As shown already in \citet{tremmel16}, their fig.9, the role of SMBH is fundamental in order to achieve a decreasing SFH with redshift as expected in  spirals of this mass \citep{Papovich15}: with feedback from stars alone, and without the contribution of SMBHs at heating the surrounding gas, such galaxies would fail at turning off their SF at late times. With the inclusion of  SMBH accretion and feedback, instead, the most massive galaxies are able to attain both a realistic stellar mass and have star formation quenched before z=0.5.
We notice that SMBH feedback  efficiently  suppresses star formation over time in a similar fashion within the two cosmologies, SIDM and CDM: the SMBHs in these galaxies have indeed similar high masses, M$_{\rm SMBH}$$>$$10^{7.7}$\msun.

In the central panels of \Fig{fig:DM-SFH} we show the evolution of DM mass within 1 kpc of the galaxy centre as a function of time. The relative contribution of adiabatic contraction and self-interactions is visible here. In the first 3 Gyrs of the galaxy life, during the rapid halo growth phase, the central DM mass increases both in the CDM and SIDM run: this phase also coincides with the first peak of star formation. Soon after the initial SF episode ends, the CDM galaxy will keep forming stars and maintaining a similar DM mass within its central region, all the way until z=0.5. On the contrary, the SIDM galaxy will undergo through a radical decrease in central DM mass, as a response to the effect of self-interactions. By z=0.5, the most massive galaxies will only keep about  30$\%$ (left panel) and  20$\%$ (right panel) of the peak inner DM mass they had at  3 Gyrs.
This corroborates the finding that the DM halo of large spiral galaxies does not follow the baryonic potential, but it is rather influenced by self-interactions  for a cross-section of $\sigma$=10 $\rm cm^2/gr$, combined with the effect of stellar and BH feedback.
Given the complexity of the physics involved, this knowledge could be only achieved by performing accurate  hydrodynamical simulations.

\begin{figure*}
\hspace{-.3cm}\includegraphics[width=3.6in,height=2.5in]{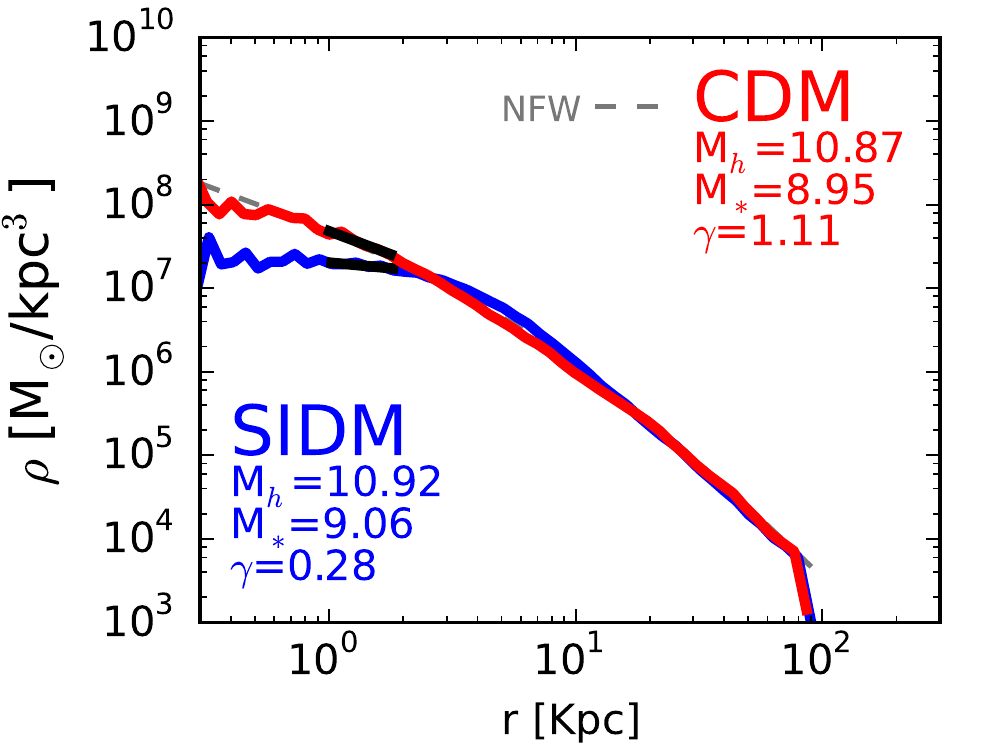}\includegraphics[width=3.6in,height=2.5in]{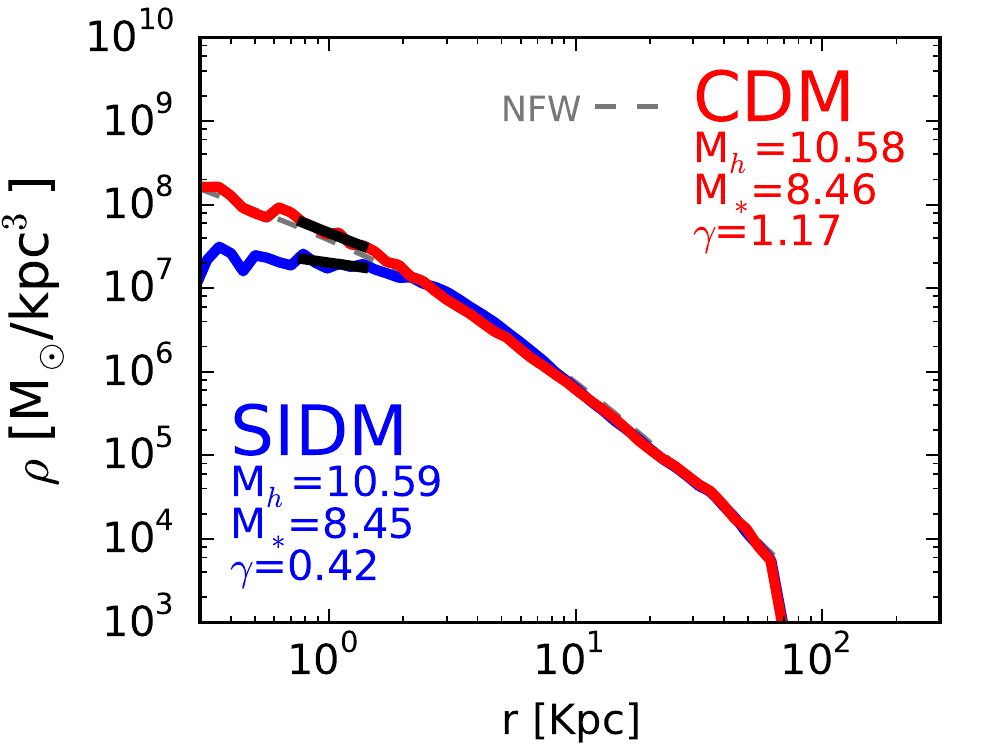}
\caption{Dark matter density profiles for two representative medium mass galaxies within the simulated SIDM and CDM volumes.  SIDM results are indicated as blue lines, CDM ones as red lines, and an unmodified NFW halo of similar mass as dashed grey line. The effect of self-interaction is evident in the large DM core formed within the SIDM galaxies, as opposed to the CDM ones that retain an initial NFW profile.}
\label{fig:intermediate-density}
  \end{figure*}

\begin{figure}
\hspace{-.3cm}\includegraphics[width=3.6in,height=4.5in]{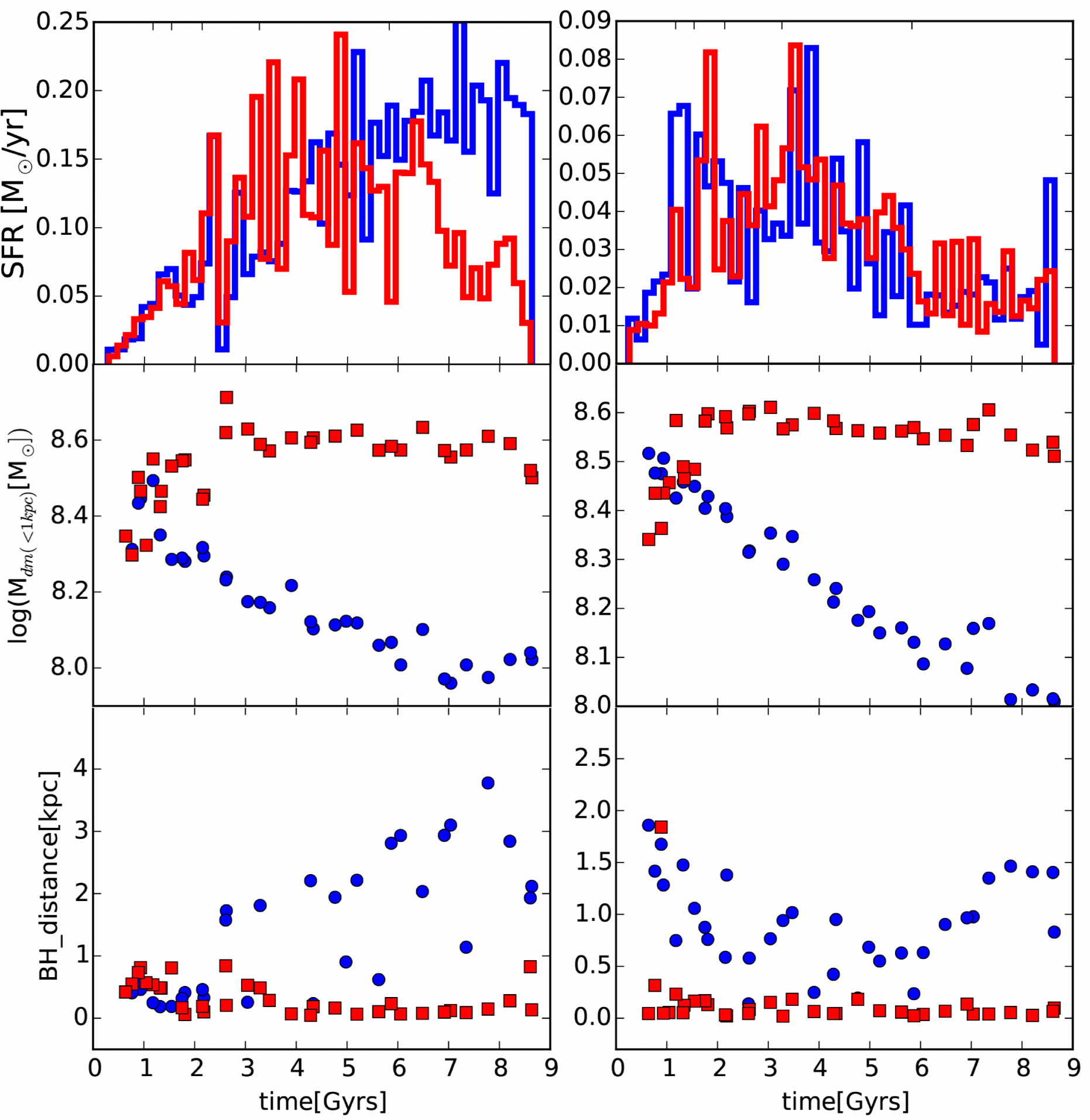}
\caption{From top to bottom: SFHs, evolution of central DM mass and evolution of SMBH distance from the galaxy centre, for the same two  medium mass  galaxies shown in \Fig{fig:intermediate-density}. Only the most massive SMBH within each galaxy is followed back in time. SIDM results are indicated as blue lines and circles, CDM ones as red lines and  squares. See text for details.}
\label{fig:dwarf-panel} 
  \end{figure}

Finally, in the bottom panels, we show the evolution of the distance of the most massive  SMBH from the center of its host galaxy, by selecting the most massive SMBH within each galaxy at  z=0.5 and following it backwards in time.
Indeed, thanks to the novel implementation of \citet{tremmel15}, we are able to accurately follow the  orbit of SMBHs through the lifetime of the galaxy and evaluate the effects of dynamical friction on the SMBH decay time. 
For the most massive galaxy, left panel, we observe that in the CDM run the galaxy undergoes  a merger at t=6.5 Gyrs that carries in the most massive SMBH. This object quickly merges with the previously existing SMBH, in a timescale of less than 0.5 Gyrs. In the SIDM run, instead, the merger that carries in the most massive SMBH happens at t=2.5 Gyrs, and it takes 2.5 Gyrs more before the SMBH can actually sink to the center of the galaxy.
For the second most massive galaxy, the SMBH goes to the center of the host by t=3-4 Gyrs, although with a longer time-scale in SIDM than CDM.
For such massive spirals, in both cosmologies the most massive SMBH  has reached the center of its host galaxy.

\subsection{Intermediate mass galaxies} \label{sec:dwarfs}
We proceed at analysing the effect of self-interactions on the haloes of two medium mass galaxies with \mhalo$\sim10^{10.5-10.9}$\msun, representative of galaxies in this mass range. The resulting dark matter profiles are shown in \Fig{fig:intermediate-density}, in blue for the SIDM case and red for the CDM one. We further show an unmodified NFW profile of same halo mass as a dashed grey line.
The halo of such medium mass galaxies shows a clear DM core with core radius of 2-5$\%$$\rm R_{vir}$ in the SIDM case, attributable to the effect of self-interactions. In the CDM run, instead, the galaxy retains the original cuspy NFW profile, with no further adiabatic contraction at work.
As expected, the effect of adiabatic contraction at these masses is minimal, and therefore the core formation due to self-interactions is the only relevant process that modifies DM haloes in SIDM galaxies.
In \Fig{fig:dwarf-panel} we show, from top to bottom, the star formation histories, temporal evolution of DM mass within 1 kpc and time evolution of the distance of the most massive SMBH from the galactic centre, for the same two galaxies of \Fig{fig:intermediate-density}.
SIDM results are shown in blue and circled points while CDM ones are shown in red  and squares.
The SFHs of the two galaxies are quite similar in the two runs, with slightly more stars formed in the SIDM case for the most massive galaxy.
The inner DM mass constantly decreases through the lifetime of the SIDM galaxies, without following the stellar potential, while it is constant in the CDM galaxies.

An interesting finding concerns the location of SMBHs within these galaxies. As  seen in the bottom panels of \Fig{fig:dwarf-panel}, the SMBH never sinks to the centre of its host galaxy in the SIDM case, while it is found  within a distance smaller than 0.1 kpc in most of the time-snapshots in CDM.
The most massive SMBHs within the galaxies in the left-hand panel have a mass of M$_{\rm SMBH}$=10$^{6.7}$ and 10$^{6.8}$\msun\ in the CDM and SIDM case, respectively, and they are found at 0.13 and 2.12 kpc from the center of their host galaxy. Similarly, the most massive SMBHs within the right-hand panel galaxy have M$_{\rm SMBH}$=10$^{6.3}$ and 10$^{6.6}$\msun\ and lie at 0.09 and 0.9 kpc from the galaxy centre, in CDM and SIDM respectively.

The fact that the SMBHs cannot reach the centre of their galaxy in the SIDM case is explained by the lower dynamical friction force that they feel: in the presence of a core, such as the one caused by self-interactions on the DM distribution, the sinking object goes through a `stalling' phase in which essentially no dynamical friction is experienced.
Such behaviour was firstly reported and analysed by \citet{read06} using analytic methods and N-body simulations, 
and we confirm here its validity using hydrodynamical simulations.
In the presence of a cored central distribution of mass, such as the one found in SIDM galaxies, we expect  SMBHs not to be found at the centre of their host galaxies.  Of course, the timescale over which each SMBH sinks depends further on its mass: this is why, in the case of the massive SMBHs belonging to MW-like galaxies of \Fig{fig:DM-density}, we found them at the centre of their hosts both in SIDM and CDM. Moreover, in the most massive galaxies, the inner slope of DM   never approaches the zero value, which is the key for having a `stalling' behaviour.

The `stalling' of SMBHs in SIDM is clearly visible in both bottom panels of \Fig{fig:dwarf-panel}. 
Intriguingly, in the left panel, we can observe the SIDM SMBH orbit oscillating, and surprisingly drifting away from the galaxy center as we move towards low redshifts.
This rather unexpected behaviour seems to correlate with the SFH of  the galaxy.  A possible explanation is  that SN-driven gas outflows, launched during SF episodes,  have a different impact on  the SIDM galaxy, whose gravitational potential in the inner region has been modified by the self-interacting particles.  If  this is the case, we should observe a correlation between gas outflows and SMBHs orbits.  As a proxy for gas outflows, we verified that the gas mass that gets ejected from the inner 1 kpc of the galaxy centre throughout its lifetime  correlates with the SMBHs orbits, such that the SMBHs drift away just after an outflow has occurred. The different extension and impact of SN-driven outflows in SIDM simulations was also argued in \cite{vogelsberger14},  simulating dwarfs of \mhalo$\sim10^{10}$\msun. 

While this finding deserves further investigation, and needs to be verified by means of higher resolution simulations, the picture that emerges is the following: strong outflows of gas are more efficient in the presence of an initial core - like the one formed in  SIDM galaxies - and are able to influence further the DM, stellar and SMBH distribution in SIDM galaxies. The SMBH responds to rapid variation in the potential in the same way as DM and stars, by moving and drifting outwards.  
In the CDM case, instead, the impact of outflows on an initially NFW halo is minimal given the low SF density threshold implemented, such that the small variations in central potential are never able to create a DM core nor to influence the dynamics of SMBHs: subsequent outflows  are simply not able to modify neither the DM, nor the stellar distribution, nor the SMBH position.

\subsection{Black holes global properties} \label{sec:SMBHs}
\begin{figure*}
\hspace{-.5cm}\includegraphics[width=3.6in,height=3.2in]{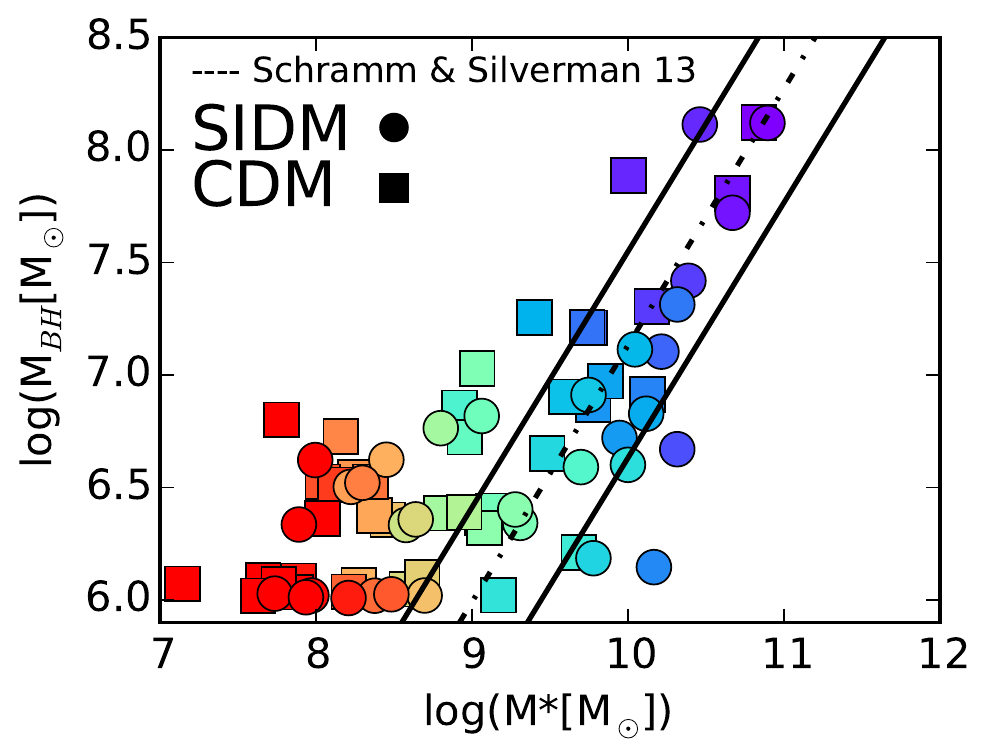}
\hspace{-.3cm}\includegraphics[width=3.6in,height=3.2in]{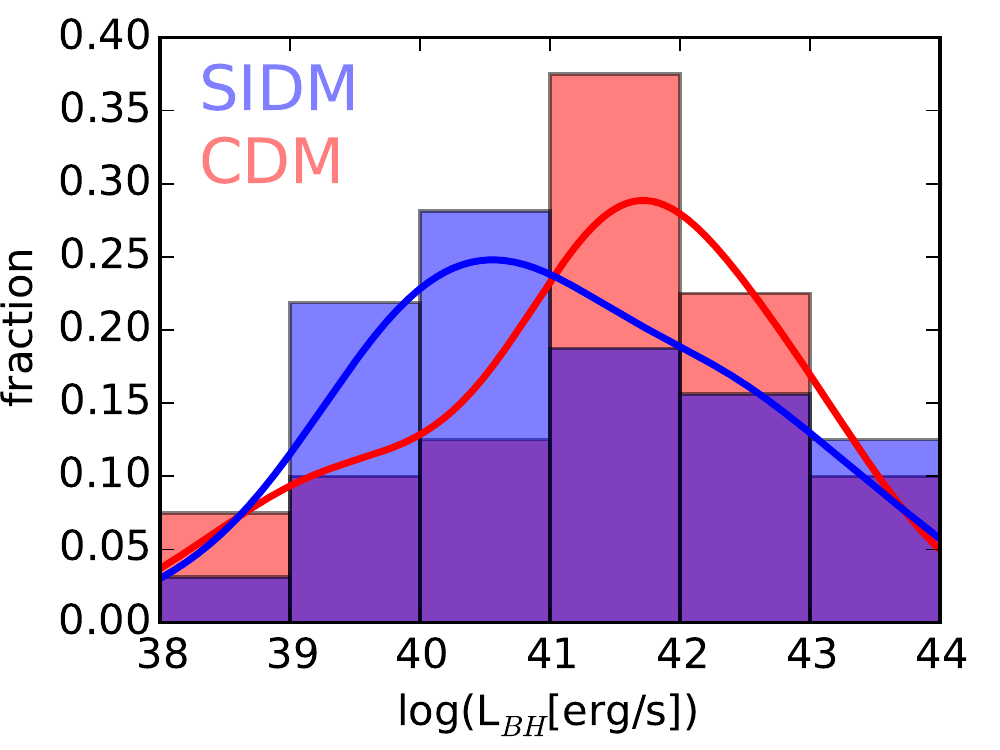}
\caption{Global properties of the SMBHs population in SIDM and CDM galaxies. Only the most massive SMBH in each galaxy is shown. \textit{Left-hand panel}: SMBH mass-host stellar mass relation, SMBHs in SIDM galaxies are indicated as circles and corresponding CDM ones as squares. Same colours refer to the same galaxy in the two runs. \textit{Right-hand panel}: Normalized histograms and PDFs of SMBH luminosities, SIDM results are shown in blue, CDM ones in red.}
\label{fig:SMBHs_distance_histograms} 
\end{figure*}
The properties of  the SMBHs population in  SIDM and CDM runs are explored in this section. We identified the most massive SMBH  associated with each central galaxy in our sample, at the latest simulated redshift  z=0.5. Galaxies less massive than  \mstar$\sim \rm 10^{7.5}$\msun\  in SIDM and  \mstar$\sim \rm10^{7.0}$\msun\  in CDM do not possess any SMBH. We have excluded galaxies with on-going mergers which are carrying in the main SMBH of a halo, since this would misleadingly provide a large distance of the SMBH from the  centre of  the galaxy within which it is just falling in. This criterion eliminates two SMBHs in the CDM run and two in the SIDM one. We have instead left in our sample galaxies with recent mergers if their most massive SMBH belongs to the same galaxy  before and after the merger itself.

We identified a total of 33 SMBHs in SIDM and 41 in CDM, each of them associated with an individual galaxy in the corresponding simulation.
In \Fig{fig:SMBHs_distance_histograms} we show, from left to right, the  SMBH mass-galaxy stellar mass relation (circles for SIDM and squares for CDM) and normalized histograms of   luminosities, with corresponding probability density functions (PDFs),  for the full SMBH population in SIDM (blue) and CDM (red) runs.  Luminosities are computed starting from accretion rates averaged over few Myrs, and taking into account  that radiative efficiency is assumed to be 0.1.  In the left-hand panel of  \Fig{fig:SMBHs_distance_histograms},  same colours refers to the same galaxy within the two runs, and overimposed as a black line is the M$_{\rm SMBH}$-\mstar\ relation of active galactic nuclei obtained with the  \textit{Chandra} deep field survey \citep{schramm13}.

The SMBH physics prescriptions adopted in this work, following \citet{tremmel15,tremmel16}, are able to reproduce a realistic relation between the SMBH mass and their host galaxy stellar mass in both SIDM and CDM cosmologies \citep{haring04,schramm13}. At low  SMBH masses, towards the lower end of the mass function (M$_{\rm SMBH}$$\sim$$10^{6}$\msun), the scatter in the relation increases as noticed already in \citet{tremmel16}. At a fixed stellar mass, several SMBHs in the CDM run tend to be more massive than the SIDM ones, reflecting the fact that SMBHs in a dense CDM halo can grow more than their SIDM counterpart. The two most massive galaxies instead  host SMBHs with a similarly high mass of M$_{\rm SMBH}$$>$$10^{7.5}$\msun.
The overall distribution of SMBH luminosities in the right-hand panel of \Fig{fig:SMBHs_distance_histograms} shows that the peak of luminosity of CDM-SMBHs happens at  higher luminosities than the corresponding SIDM case: from the PDFs we estimated a peak at $\rm L_{SMBH}$$=$$10^{41.7}$erg/s in the CDM case, and at  $\rm L_{SMBH}$$=$$10^{40.5}$erg/s in the SIDM case. While, of course, caution should be exercised due to the low number statistic, the reason  for this has to be searched  in the SMBHs evolution and accretion: SMBHs forming within galaxies with a  shallow central profile, like the  ones of SIDM haloes, will accrete less mass and therefore have a lower luminosity than the corresponding  CDM case, in which SMBHs accrete in a  denser environment with higher average gas supply that can reach into the SMBH.

\vspace{-.2cm}
\subsection{Black holes dynamics} \label{sec:SMBHs_dyn}
  \begin{figure*}
\includegraphics[width=7.in,height=3.2in]{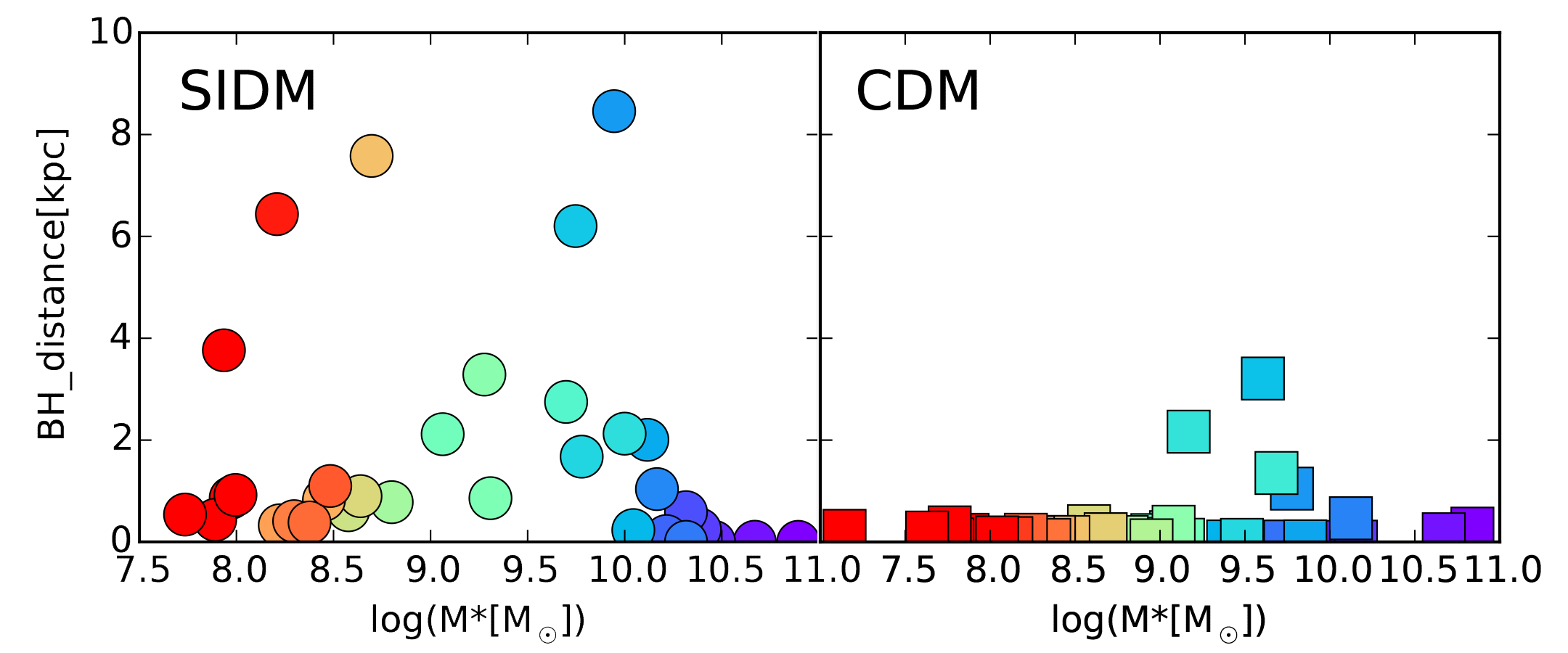}\caption{SMBH distance from the centre of its host galaxy as a function of host stellar mass, in SIDM (\textit{left-hand panel}, circles) and CDM (\textit{right-hand panel}, squares), respectively. Same  colours refer to the same galaxy in the two runs. Only the most massive SMBH within each galaxy is shown.}
\label{fig:SMBHs_distance} 
\end{figure*}
Most interestingly, we found that the dynamics of SMBHs is different in the two cosmologies and that, in the SIDM case, it has been affected by the modification to the halo structure due to self-interactions between DM particles.
In \Fig{fig:SMBHs_distance} we plot the distance of SMBHs from host centre vs host stellar mass, with SIDM results indicated as circles (left-hand panel) and CDM ones as squares (right-hand panel).  
As in the previous section, only the most massive SMBH within each galaxy is shown, and same colours refer to the same galaxy in the two cosmologies.

It appears that several SMBHs in SIDM galaxies have not  yet reached the galaxy centre, being found as far as $\sim$9 kpc from it, while the vast majority of SMBHs in CDM lie within 1 kpc from the centre of the host galaxy.
We observe  a striking difference in the behaviour of SMBHs in SIDM and CDM, already anticipated in \Sec{sec:dwarfs}: SMBHs in the SIDM run are on average further away from the galaxy center than their CDM counterparts. Only 4  outlier SMBHs lie at or above 1 kpc from the host's center in CDM galaxies, while in the case of SIDM we found as many as 13 SMBHs above such a distance. In the SIDM case SMBHs are found as far as $\sim$9 kpc from the host center.
More specifically, 88$\%$ of all the SMBHs identified in the CDM simulation have sunk to the center of their host galaxy and lie within 350 pc (corresponding to the resolution limit of our simulation, $\epsilon$) from it by z=0.5, while only about  20$\%$ of the  SMBHs in the SIDM run lie within the same distance from their host galaxy.
All the  SMBHs that lie within $\epsilon$ from the center of SIDM galaxies are found in galaxies more massive than \mstar=$10^{10}$\msun, indicating that the mechanism responsible for keeping SMBHs away from the centre  is more dramatic in lower mass galaxies. 
Recall  that our SMBH model is able to effectively take into account the dynamical friction forces acting on the SMBHs, which are therefore \textit{not} forced at the center of the galaxy as in previous prescriptions  \citep{dimatteo05,sijacki07}  but  they are  rather allowed to orbit and sink at the center of their hosts over realistic timescales. 

We can therefore understand the trends shown in \Fig{fig:SMBHs_distance} in terms of dynamical friction timescales \citep{chandrasekhar43,BT08,read06}. 
Because the timescale  is inversely proportional to the density of the surrounding medium,  SMBHs sinking within lower density SIDM haloes have not yet reached the center of their host, and are `stalling' within the core of SIDM galaxies, in agreement with the analytic model presented in \citet{read06} and \citet{petts16}. The `stalling' is due to the failing of the Chandrasekar formula in the presence of a central density core, and reflects the situation in which SMBHs effectively do not feel any dynamical friction.  In comparison, a much more centrally concentrated CDM halo will have a short enough dynamical friction timescale for the SMBHs to have sunk all the way to its center.
Moreover, as already investigated in \Sec{sec:dwarfs}, the increased efficiency of gas outflows in the presence of an underlying cored distribution further contributes to keep the SMBH away from the galactic centre during vigorous SF episodes. Both effects are clearly shown in \Fig{fig:dwarf-panel},  by comparing  a SIDM and CDM halo of similar masses. 
The  SMBHs found within the highest stellar mass galaxies (\mstar$>10^{10.5}$\msun) have instead reached the center of their hosts in both SIDM and CDM, as shown explicitly in \Fig{fig:DM-SFH}: this is because despite of the lower central density of such galaxies in the SIDM run, their associated SMBHs are  massive enough to  dominate  the dynamical friction timescale, allowing to sink all the way to the host's  centre.
The dramatic effect of longer dynamical friction timescales observed in SMBHs within SIDM simulations, combined with the increased impact of gas outflows on the central potential of SIDM galaxies, implies that the majority of galaxies with masses below \mstar$\sim10^{10.5}$\msun\ are expected to have a SMBH offset from their center in a SIDM universe.

\vspace{-.2cm}\section{Conclusions}\label{sec:concl}
We investigated the properties of galaxies in hydrodynamical cosmological simulations, run in a 8Mpc box down to redshift z=0.5, within a cold dark matter (CDM) as well as a self-interacting dark matter (SIDM) scenario using same initial conditions. The implementation of SIDM  follows the already published work of \citet{bastidas15}. We used a constant cross-section of $\sigma$=10 $\rm cm^2/gr$, which is close to the upper limit for Milky Way like galaxies \citep{kaplinghat15}, to maximise the effect of self-interactions on dark matter profiles.
The simulations, run with the N-body + SPH code ChaNGA \citep{menon15}, include star formation, UV background, blast wave feedback from SN, thermal diffusion \citep{wadsley04,wadsley08,stinson06,shen10}, as well as an improved prescription for SMBH  physics as in \cite{tremmel15}. 
SMBHs form from dense, low metallicity gas at early times, accrete gas according to a modified Bondi-Hoyle formula that accounts for the rotational support of gas, and feel a realistic dynamical evolution through  dynamical friction prescription \citep{tremmel15,tremmel16}, allowing SMBHs to experience realistic perturbations and sinking timescales during satellite accretion and galaxy merger events. 
The prescription for SMBH growth and feedback has been calibrated, along with star formation and feedback processes, to produce galaxies with realistic stellar and SMBH masses \citep{tremmel16}.
The star formation threshold of $\rm0.2m_p/cm^3$ used in both cosmologies  is too low to lead to the creation of central dark matter cores via outflows \citep{Pontzen12}. 
In contrast, with such a high cross-section $\sigma$, DM cores are expected to form in pure SIDM haloes even at the lowest masses considered here, \mhalo$\sim$$10^{10}$\msun\ (see \citealt{rocha13} and \citealt{bastidas15} for similar discussion).
The main results of this work can be summarized as follows:\\

\noindent\textbf{Density profiles and SFHs}\begin{itemize}
\item  massive galaxies, \mhalo$\sim$$\rm10^{12}$\msun, show a less dense dark matter profile in the SIDM case compared to the CDM one,  with the SIDM halo density falling below the CDM one already at radii of 20 kpc, being the effect of self-interactions dominant over the adiabatic contraction. This is despite  the similarly dominating baryonic potential at the centre of such massive galaxies in both cosmologies. Feedback from the central SMBH is efficient at regulating star formation in both the SIDM and CDM runs \citep{tremmel15}, and the most massive galaxies show a similarly declining SFH, reminiscent of the one observed in massive spirals;
\item medium mass  galaxies, $10^{10}$$<$\mhalo/\msun$<$$10^{11}$, show a well defined DM core in the SIDM run, while a usual NFW halo in the CDM one. The effects of both SMBH feedback and adiabatic contraction are not dominant at these scales, and the only  modification to the DM profile is due to self-interactions (see also \citealt{zavala13,vogelsberger12,bastidas15});
\end{itemize}

\noindent\textbf{Black hole dynamics}
\begin{itemize}
\item the CDM run produces more high-luminosity SMBHs with respect to the SIDM case ($\sim$70$\%$ of CDM SMBHs have $\rm L_{SMBH}$$>$$10^{41}$erg/s, while only 45$\%$ of the SIDM ones are found above this value), a consequence of the higher accretion rate due to the denser local environment around the SMBHs;
\item the dynamics of SMBHs changes in the two runs. In the SIDM case, due to the longer dynamical friction timescales \citep{chandrasekhar43,taffoni03} caused by the lower density in the central regions of dark matter haloes, the SMBHs experience the phenomenon of \textit{core stalling}, in agreement with analytic predictions from \citet{read06}, and never reach the host galaxy centre. SMBHs are found at a distance of up to 9 kpc from the galaxy centre in SIDM. Only about 20$\%$ of the most massive SMBHs associated with SIDM galaxies have reached a distance from the centre smaller than our resolution limit, 350 pc, while as many as $\sim$90$\%$ of the SMBHs in CDM are found well within this value.
\end{itemize}
In a  SIDM cosmology with a large constant cross section of $\sigma$=10 $\rm cm^2/gr$, about 83$\%$ of the galaxies with masses lower than \mstar$\sim$$10^{10.5}$\msun\ should have a SMBH offset from their center due to the effect of longer dynamical friction timescales caused by shallow density cores.
This prediction could lead to a series of potentially observable quantities, such as anomalies in the stellar velocity dispersion at the SMBH location, and further activities connected to gas accretion, such as off-center bright nuclei and off-center X and radio sources. \\

\noindent Our study highligths the critical importance of  properly modeling baryonic physics processes and SMBH dynamics within different underlying DM models. We plan to extend this work to higher resolution simulations in which baryonic driven core formation will  play a role as well. This will help verifying whether  the offset of SMBHs occurs even in galaxies whose cores are generated by stellar feedback, rather than by self-interactions.

\vspace{-.5cm}
\section*{Acknowledgements}
All simulation analysis made use of the \textit{pynbody} \citep{Pontzen13}  and TANGOS (Pontzen  et  al.  in  prep)  suites. ADC is supported by the DARK-Carlsberg and  the Karl Schwarzschild independent fellowship programs. ADC thanks the Mainz Institute for Theoretical Physics (MITP) for its hospitality. MT was partially supported by NSF award AST-1514868. FG acknowledges support from NSF grant AST-0607819 and NASA ATP NNX08AG84G. AMB acknowledge support from NSF grant AST-1411399. MV acknowledges support through an MIT RSC award and the support of the Alfred P. Sloan Foundation.
\vspace{-.2cm}
\bibliographystyle{mn2e}
\bibliography{archive}


\label{lastpage}

\end{document}